\documentclass[preprintnumbers,article,amsmath,amssymb,floatfix,10pt,prd,onecolumn,superscriptaddress,nofootinbib]{revtex4-2}

\usepackage{titlesec}
\usepackage[toc]{appendix}
\usepackage{caption}
\titleformat*{\section}{\LARGE\bfseries}
\titleformat*{\subsection}{\Large\bfseries}
\titleformat*{\subsubsection}{\large\bfseries}
\titleformat*{\paragraph}{\large\bfseries}
\titleformat*{\subparagraph}{\large\bfseries}
\usepackage{bm}
\usepackage{amsfonts}
\usepackage{latexsym}
\usepackage[utf8]{inputenc}
\usepackage{graphicx}
\usepackage{amsmath}
\usepackage{palatino}
\usepackage{mathpazo}
\usepackage{textcomp}
\usepackage{float}
\usepackage{booktabs}
\usepackage{dcolumn}
\usepackage{ragged2e}
\usepackage{hyperref}
\hypersetup{colorlinks,citecolor=blue}
\hypersetup{colorlinks=true,linkcolor=blue,filecolor=magenta,urlcolor=blue}
\usepackage{amsthm}
\usepackage{xcolor}
\usepackage{orcidlink}
\usepackage{epsfig}
\usepackage{commath}
\usepackage{cancel, ulem}
\usepackage{subcaption}
\usepackage{caption}
\usepackage{multirow}
\newcommand{\be}{\begin{equation}}
\newcommand{\ee}{\end{equation}}
\newcommand{\bea}{\begin{eqnarray}}
\newcommand{\eea}{\end{eqnarray}}
\newcommand{\eeas}{\end{eqnarray*}}
\newcommand{\beas}{\begin{eqnarray*}}

\def\jnl@style{\it}
\def\aaref@jnl#1{{\jnl@style#1}}

\def\aaref@jnl#1{{\jnl@style#1}}

\def\aj{\aaref@jnl{AJ}}                   
\def\apj{\aaref@jnl{ApJ}}                 
\def\apjl{\aaref@jnl{ApJ}}                
\def\apjs{\aaref@jnl{ApJS}}               
\def\apss{\aaref@jnl{Ap\&SS}}             
\def\aap{\aaref@jnl{A\&A}}                
\def\aapr{\aaref@jnl{A\&A~Rev.}}          
\def\aaps{\aaref@jnl{A\&AS}}              
\def\mnras{\aaref@jnl{Mon.~Not.~Roy.~Astron.~Soc.}}             
\def\prd{\aaref@jnl{Phys.~Rev.~D}}        
\def\prc{\aaref@jnl{Phys.~Rev.~C}}  
\def\prl{\aaref@jnl{Phys.~Rev.~Lett.}}    
\def\qjras{\aaref@jnl{QJRAS}}             
\def\skytel{\aaref@jnl{S\&T}}             
\def\ssr{\aaref@jnl{Space~Sci.~Rev.}}     
\def\zap{\aaref@jnl{ZAp}}                 
\def\nat{\aaref@jnl{Nature}}              
\def\aplett{\aaref@jnl{Astrophys.~Lett.}} 
\def\apspr{\aaref@jnl{Astrophys.~Space~Phys.~Res.}} 
\def\physrep{\aaref@jnl{Phys.~Rep.}}      
\def\physscr{\aaref@jnl{Phys.~Scr}}       
\def\commat{\aaref@jnl{Comm.~Math.~Phys.}}              
\def\science{\aaref@jnl{Science}}               
\def\cqg{\aaref@jnl{Classical Quant.~Grav.}}            
\def\jpcs{\aaref@jnl{JPCS}}                                     
\def\ijmpd{\aaref@jnl{Int.~J.~Mod.~Phys.~D}}                    
\def\grg{\aaref@jnl{Gen.~Relat.~Gravit.}}               
\def\rpp{\aaref@jnl{Rep.~Prog.~Phys.}}          
\def\npa{\aaref@jnl{Nucl.~Phys.~A}}        
\def\lrr{\aaref@jnl{Living Rev.~Rel.}}                   
\def\jcap{\aaref@jnl{J.~Cosmology Astropart.~Phys.}}    
\def\rmp{\aaref@jnl{Rev.~Mod.~Phys.}}   
\def\epjc{\aaref@jnl{Eur.~Phys.~J.~C}} 
\def\plb{\aaref@jnl{~Phy.~Lett.~B}} 
\def\mpla{\aaref@jnl{Mod.~Phy.~Lett.~A}} 
\def\arxiv{\aaref@jnl{arxiv.org}}

\allowdisplaybreaks[1]

\begin{document}
\title{Exact cosmological solutions in non-coincidence $f(Q)$-theory}
\author{Avik De\orcidlink{0000-0001-6475-3085}}
\email{avikde@um.edu.my}
\affiliation{Institute of Mathematical Sciences, Faculty of Science, Universiti Malaya, 50603 Kuala Lumpur, Malaysia}
\author{Andronikos Paliathanasis\orcidlink{0000-0002-9966-5517}}
\email{anpaliat@phys.uoa.gr}
\affiliation{Department of Mathematics, Faculty of Applied Sciences, Durban University of
Technology, Durban 4000, South Africa}
\affiliation{Departamento de Matem\`{a}ticas, Universidad Cat\`{o}lica del Norte, Avda.
Angamos 0610, Casilla 1280 Antofagasta, Chile}
\affiliation{National Institute for Theoretical and Computational Sciences (NITheCS), South Africa}

\footnotetext{Part of this study was supported by FONDECYT 1240514, ETAPA 2025.}

\begin{abstract}
We study exact cosmological solutions in $f(Q)$ gravity formulated beyond the coincident gauge, focusing on the non-coincident connection branch $\Gamma_B$. Using a minisuperspace approach, the field equations are recast into an equivalent scalar-tensor form, enabling analytic reconstruction of cosmological models. We obtain exact solutions of particular interest, including de Sitter, scaling, $\Lambda$CDM, Chaplygin gas, generalized Chaplygin gas, and CPL parameterizations. The corresponding scalar potentials and $f(Q)$ functions are derived in closed or parametric form. Our analysis shows that non-coincident $f(Q)$ gravity admits a richer solution space than the coincident case and can describe both early-time inflationary dynamics and late-time acceleration within a unified framework. These results open new directions for testing $f(Q)$ cosmology against observations and exploring its role as a viable alternative to $\Lambda$CDM.
\end{abstract}

\maketitle

\section{Introduction}\label{sec1}

Recent observational results have confirmed that our universe has experienced at least two distinct accelerated phases of expansion: an early epoch of inflation and the present phase of late-time acceleration \cite{Riess,Tegmark,Komatsu}. These phenomena suggest that General Relativity (GR), while remarkably successful, may not be the final theory of gravitation, motivating the exploration of alternative frameworks. One promising direction is provided by the so-called ``geometric trinity of gravity,'' where torsion, curvature, or non-metricity can serve as fundamental carriers of gravitational interaction. In particular, the symmetric teleparallel framework, based on non-metricity, has attracted increasing attention, and the family of modified gravity models known as $f(Q)$ gravity has been extensively studied, elaborative details can be found in the important review \cite{surveyfQ} and the references therein.

Within non-metricity gravity, most cosmological investigations have relied on the coincident gauge, where the connection coefficients are zero. This although simplifies the system, the field equation reduces to that of the corresponding metric teleparallel counterpart, precisely, the $f(T)$ gravity \cite{surveyfT}. Recent developments have emphasized the importance of exploring the general non-coincident connections, where additional degrees of freedom may leave imprints on cosmological dynamics. Such an approach allows for a richer solution space and could reveal hidden sectors of phenomenology not visible in the coincident branch. The dynamical system analysis of $f(Q)$ gravity from non-coincident branches was carried out in \cite{Paliathanasis:2023nkb,shabanifQ,Narawade:2024pxb,ghulamfQ,saikatfQ} and a true sequence of cosmic eras was demonstrated. Recently, in \cite{rajadata} the authors investigated a very specific form of a non-coincident branch from a Hubble parameterization. By employing Hubble and Gaussian processes, a data reconstruction of the dynamical degree of freedom in non-coincident branches were carried out for two of the most studied $f(Q)$ gravity models \cite{saridakisdata}. At the background level, $\Lambda$CDM mimicking $f(Q)$ gravity formulated from non-coincident class of connections were investigated in \cite{saikatfQlcdm}, analytic reconstruction was made possible for connection class II, and numerical reconstruction for class III using a cosmographic condition.  

On the other hand, the non-coincident formulation of power-law $f(Q)$ gravity was shown to challenge $\Lambda$CDM from DESI DR2 \cite{an01}. The consideration of the non-coincident formulation in $f(Q)$ leads to the introduction of new dynamical degrees of freedom that can be attributed to scalar fields. The latter can describe cosmic acceleration without the introduction of the cosmological constant term, or another matter component \cite{Paliathanasis:2023pqp}. An interesting discussion on the importancy of the non-coincidence gauge presented recently in \cite{Ayuso:2025vkc}

It is worth emphasizing that certain instabilities in $f(Q)$ gravity have been identified in \cite{R6}, where cosmological perturbations display pathological features. Among the three spatially flat cosmological branches, two are plagued by infinitely strong coupling, rendering their linear spectra physically non-viable, while the remaining branch propagates seven gravitational degrees of freedom, including at least one ghost excitation, thereby threatening theoretical consistency. Furthermore, \cite{R7} argued that a scalar mode in the $f(Q)$ framework unavoidably carries negative kinetic energy, indicating a ghost instability, irrespective of the commonly adopted coincident gauge. Nevertheless, by reformulating the theory within a higher-order scalar–tensor representation and substituting scalar St\"uckelberg fields with vector fields, it was shown that the second-class constraints in the Arnowitt–Deser–Misner formalism can eliminate this ghost degree of freedom.

Exact solutions play a central role in cosmology, since they provide a laboratory to test theoretical consistency, probe singularity structure, and compare with observational data without resorting solely to numerical methods. In General Relativity, a wealth of exact solutions is known for scalar-field cosmologies, Chaplygin gas universes, and anisotropic Bianchi models. Similarly, in modified gravity, exact or closed-form solutions have been extensively studied in \(f(R)\) and \(f(T)\) frameworks. For instance, in \(f(T)\) gravity one can obtain power-law solutions for FLRW cosmologies, de Sitter solutions relevant for inflation and dark energy, as well as anisotropic Bianchi-type exact solutions which provide insights into isotropization and early-universe dynamics \cite{Wu2010,Ferraro2007,Paliathanasis:2016vsw,Oikonomou:2025xms,Oikonomou:2023env,Luongo:2024opv,DAgostino:2021vvv,Carloni:2024ybx}.

More recently, exact cosmological and astrophysical solutions have begun to emerge within \(f(Q)\) gravity itself. Power-law models such as \(f(Q)=Q+\alpha Q^n\) have been shown to admit analytic cosmological solutions that effectively mimic \(\Lambda\)CDM behavior at both background and perturbation levels; integrability is sometimes demonstrated using methods like the Painlevé test \cite{Khyllep2021}. Furthermore, topological and spherically symmetric vacuum or wormhole solutions have been obtained, including static black hole or wormhole geometries in models where the non-metricity scalar \(Q\) is constant or in power-law forms, demonstrating that \(f(Q)\) supports rich exact solutions beyond the coincident FLRW setting \cite{Dimakis:2024fan,Celclass2023,Alwan:2025nka,Alwan:2024lng}.

Beyond providing explicit metrics, exact solutions often emerge from integrability techniques, including the use of minisuperspace Lagrangians, Noether symmetries, and dynamical system analysis \cite{Tsamparlis:2018nyo,Dimakis:2022pks,Dimakis:2021rgz,Terzis:2014cra,Dialektopoulos:2021ryi,Dialektopoulos:2025ihc,Dialektopoulos:2021ryi,Cid:2017wtf,Leon:2021wyx,Leon:2012vt,Gonzalez:2006cj,Tripathy:2024per}. Such approaches allow one to classify solution families systematically, identify attractors of cosmological evolution, and reveal the role of conserved quantities. In this sense, exact solutions not only illustrate specific cosmological models but also act as cornerstones for understanding the broader phase space of modern cosmology and modified gravity theories. We refer the reader to the interesting discussion on the importancy of exact solutions in \cite{MacCallum:1984zxc}.

In contrast, analytic cosmological solutions in non-coincident \(f(Q)\) gravity remain largely uncharted territory. The non-coincident formulation admits additional structure via nontrivial affine connections that could play a decisive role in the dynamics of both the early and late universe. The aim of this paper is to fill this gap by investigating exact cosmological solutions in the framework of non-coincident \(f(Q)\) gravity. By employing a minisuperspace approach in the context of the non-coincident connection and using dynamical system or symmetry techniques, we derive families of exact solutions for homogeneous and isotropic cosmologies, and discuss their physical implications. In particular, we demonstrate that these exact solutions may reproduce well-known inflationary and late-time accelerating behaviors, while also admitting new features absent in the coincident case. Our analysis thus provides a step toward a more complete understanding of the cosmological potential of \(f(Q)\) gravity beyond the coincident gauge. The structure of the paper is as follows.

In Section \ref{sec2} we introduce the symmetric teleparallel $f(Q)$-gravity. We focus in the case of a spatially flat FLRW geometry in
which the symmetric and flat connection is defined in the non-coincidence
gauge. We present the field equations in the equivalent form of scalar field
description. Section \ref{sec3} includes the main results of this study where
we investigate the existence of analytic cosmological solutions of special
interests. We make use of the scalar field description and we reproduce
previous results for the de Sitter and the self-similar solutions. However, we
show that $f\left(  Q\right)  $-gravity can describe and other solutions of
special interests for the description of inflation, as the Chaplygin gas
solutions. Moreover, we consider the case of effective parametric dark energy
models which are used for the study of the late-time cosmological
observations. Finally, in Section \ref{sec4} we draw our conclusions.

\section{$f\left(  Q\right)  $-gravity fundamentals}

\label{sec2}

In symmetric teleparallel $f\left(  Q\right)  $-gravity the fundamental
geometric object is the nonmetricity scalar $Q$ defined by a symmetric and
flat connection different from the Levi-Civita connection, where the
gravitational Action Integral is defined as \cite{Koivisto2,Koivisto3}%
\begin{equation}
S_{f\left(  Q\right)  }=\int d^{4}x\sqrt{-g}f\left(  Q\right)  . \label{eq.01}%
\end{equation}
where $f\left(  Q\right)  $ is a smooth differentiable function.

In symmetric teleparallel theory, the geometry which describes the physical
world is embedded with a metric tensor $g_{\mu\nu}$ and the connection
$\Gamma_{\mu\nu}^{\kappa}$, which has the properties, it is flat, from where
we infer that the Riemann tensor has zero components
\begin{equation}
R_{\;\lambda\mu\nu}^{\kappa}\left(  \Gamma\right)  =\frac{\partial
\Gamma_{\;\lambda\nu}^{\kappa}}{\partial x^{\mu}}-\frac{\partial
\Gamma_{\;\lambda\mu}^{\kappa}}{\partial x^{\nu}}+\Gamma_{\;\lambda\nu
}^{\sigma}\Gamma_{\;\mu\sigma}^{\kappa}-\Gamma_{\;\lambda\mu}^{\sigma}%
\Gamma_{\;\mu\sigma}^{\kappa}, \label{eq.02}%
\end{equation}
and symmetric, which follows that the torsion tensor has also zero components%
\begin{equation}
\mathrm{T}_{\mu\nu}^{\kappa}\left(  \Gamma\right)  =\frac{1}{2}\left(
\Gamma_{\;\ \mu\nu}^{\kappa}-\Gamma_{\;\ \nu\mu}^{\kappa}\right)  .
\label{eq.03}%
\end{equation}

Consequently, only the nonmetricity $Q_{\kappa\mu\nu}=\nabla_{\kappa}g_{\mu
\nu}$ contributes to the gravitational field. The nonmetricity scalar $Q$ is
defined as
\begin{equation}
Q=Q_{\kappa\mu\nu}P^{\kappa\mu\nu} \label{eq.04}%
\end{equation}
where $P_{\;\mu\nu}^{\kappa}$ is defined as%
\begin{equation}
P_{\;\mu\nu}^{\kappa}=\frac{1}{4}Q_{\;\mu\nu}^{\kappa}+\frac{1}{2}%
Q_{(\mu\phantom{\lambda}\nu)}^{\phantom{(\mu}\kappa\phantom{\nu)}}+\frac{1}%
{4}\left(  Q^{\kappa}-\tilde{Q}^{\kappa}\right)  g_{\mu\nu}-\frac{1}{4}%
\delta_{\;(\mu}^{\kappa}Q_{\nu)} \label{eq.05}%
\end{equation}
where $Q_{\mu}=Q_{\mu\nu}^{\phantom{\mu\nu}\nu}~,~\tilde{Q}_{\mu
}=Q_{\phantom{\nu}\mu\nu}^{\nu\phantom{\mu}\phantom{\mu}}$.

Let $\hat{\Gamma}_{\mu\nu}^{\kappa}$ be the Levi-Civita connection for the
metric tensor, that is, $\hat{\Gamma}_{\mu\nu}^{\kappa}=\frac{1}{2}%
g^{\kappa\lambda}\left(  g_{\mu\kappa,\nu}+g_{\lambda\nu,\mu}-g_{\mu
\nu,\lambda}\right)  $, and corresponding curvature tensor $\hat{R}%
_{\;\lambda\mu\nu}^{\kappa}\left(  \hat{\Gamma}\right)  $ and Ricci scalar
$\hat{R}$. Then, the nonmetricity scalar $Q$ for the connection $\Gamma
_{\;\ \mu\nu}^{\kappa}$ is related to $\hat{R}$ by a boundary term, that is
$Q=\hat{R}+B$, in which $B=-\frac{1}{2}\mathring{\nabla}_{\lambda}P^{\lambda}$. Consequently, when $f\left(  Q\right)  $ is a linear function,
then the gravitational Action Integral (\ref{eq.01}) describes the STEGR which is a gravitational theory equivalent to the
GR~\cite{trinity}. Hence, in the following we focus in the case of nonlinear
functions $f\left(  Q\right)  $.

\section{Cosmological aspects of $f(Q)$ gravity}

We consider a isotropic and homogeneous spatially flat FLRW geometry described
by the line element%
\begin{equation}
ds^{2}=-N^{2}\left(  t\right)  dt^{2}+a^{2}\left(  t\right)  \left(
dx^{2}+dy^{2}+dz^{2}\right)  , \label{eq.06}%
\end{equation}
where $a\left(  t\right)  $ is the scale factor describes the radius of the
universe and $N\left(  t\right)  $ is the lapse function. For the comoving
observer $u^{\mu}=\frac{1}{N\left(  t\right)  }\delta_{t}^{\mu}$, $\ u^{\mu
}u_{\mu}=-1$, the Hubble function which describes the expansion of the
universe is defined as $H=\frac{1}{N}\frac{\dot{a}}{a}$, where a dot means
total derivative with respect to the time, that is, $\dot{a}=\frac{da}{dt}$.

The definition of the connection is essential for the $f\left(  Q\right)
$-gravity. For the line element (\ref{eq.06}) the requirements the connection
to be flat, symmetric and inherits the symmetries of the FLRW spacetime leads
to three different families of connections~\cite{Hohmann,fq4,ndim}, namely
$\Gamma^{A}$, $\Gamma^{B}$ and $\Gamma^{C}$. Connection $\Gamma^{A}$ is
defined in the coincidence gauge and the cosmological field equations are
equivalent to the $f\left(  T\right)  $ teleparallel gravity. On the other
hand, connections $\Gamma^{B}$ and $\Gamma^{C}$ are defined in the
non-coincidence gauge and the cosmological field equations can be recast in form of 
multi-scalar field theories.

In this study we focus in the connection $\Gamma^{B}$ defined in the
non-coincidence gauge. Recently, it was found by using the late-time
cosmological observations that the power-law $f\left(  Q\right)  \simeq
Q^{\frac{n}{n-1}}$ cosmological model within the connection $\Gamma^{B}$
challenge the $\Lambda$CDM theory \cite{an01}. The analytic solutions for this
power-law model was determined before in \cite{Dialektopoulos:2025ihc} with the use of Noether
symmetry analysis.

We are interested in the viability of other analytic cosmological solutions
investigated in modern cosmology. In particular we prove the existence of
solutions of special interests for the cosmological evolution and we
reconstruct analytically or numerically the corresponding $f\left(  Q\right)
$ functions. 


For connection $\Gamma^{B}$ the nonzero components of the connection are%
\[
\Gamma_{\;tt}^{t}=\frac{\ddot{\psi}(t)}{\dot{\psi}(t)}+\dot{\psi}%
(t),~\Gamma_{tx}^{x}=\Gamma_{ty}^{y}=\Gamma_{tz}^{z}=\dot{\psi}\left(
t\right)  .
\]
where scalar field $\psi\left(  t\right)  $ describes the geometrodynamical
degrees of freedom introduced in the gravitational field by the connection.
For this connection, the nonmetricity scalar is derived%
\begin{equation}
Q=-6H^{2}+\frac{3\dot{\psi}}{N}\left(  3H-\frac{\dot{N}}{N^{2}}\right)
+\frac{3\ddot{\psi}}{N^{2}}.\label{eq.07}%
\end{equation}
and the the cosmological field equations of $f\left(  Q\right)  $-gravity are%
\begin{gather}
3H^{2}f^{\prime}+\frac{1}{2}\left(  f(Q)-Qf_{,Q}\left(  Q\right)  \right)
+\frac{3\dot{\psi}\dot{Q}f^{\prime\prime}}{2N^{2}}=0,\label{eq.08}\\
-\frac{2\left(  f^{\prime}H\right)  ^{\cdot}}{N}-3H^{2}f^{\prime}%
-\frac{\left(  f(Q)-Qf_{,Q}\left(  Q\right)  \right)  }{2}+\frac{3\dot{\psi
}\dot{Q}f^{\prime\prime}}{2N^{2}}=0,\label{eq.09}\\
\dot{Q}^{2}f_{,QQQ}+\left[  \ddot{Q}+\dot{Q}\left(  3NH-\frac{\dot{N}}%
{N}\right)  \right]  f_{,QQ}=0.\label{eq.10}%
\end{gather}
where $f_{,Q}\left(  Q\right)  =\frac{df}{dQ}$. Equations (\ref{eq.08}),
(\ref{eq.09}) are the modified Friedmann equations, while equation
(\ref{eq.10}) define the equation of motion for the connection, that is,
scalar $\psi$.

We introduce the scalar field $\phi=f^{\prime}\left(  Q\right)  $, and the
potential function $V\left(  \phi\right)  =\left(  f(Q)-Qf^{\prime}\left(
Q\right)  \right)  $ such that $Q=-V_{,\phi}$.\ Then the latter field
equations are expressed in the equivalent form of a multi-scalar field
cosmological model. Indeed, the modified Friedmann equations (\ref{eq.08}),
(\ref{eq.09}) read~\cite{Paliathanasis:2023pqp}%
\begin{align}
3\phi H^{2}+\frac{3}{2N^{2}}\dot{\phi}\dot{\psi}+\frac{V\left(  \phi\right)
}{2} &  =0,\label{eq.11}\\
-\frac{2}{N}\left(  \phi H\right)  ^{\cdot}-3\phi H^{2}+\frac{3}{2N^{2}}%
\dot{\phi}\dot{\psi}-\frac{V\left(  \phi\right)  }{2} &  =0,\label{eq.12}%
\end{align}
while the equation of motion for the connection becomes%
\begin{equation}
\frac{1}{N}\left(  \frac{1}{N}\dot{\phi}\right)  ^{\cdot}+\frac{3}{N}%
H\dot{\phi}=0.\label{eq.13}%
\end{equation}
Finally, the the nonmetricity scalar (\ref{eq.07}) is terms of the scalar
field description is as follows
\begin{equation}
V_{,\phi}=6H^{2}-\frac{3}{N}\left(  \dot{\psi}\left(  3H-\frac{\dot{N}}{N^{2}%
}\right)  +\frac{1}{N}\ddot{\psi}\right)  .\label{eq.14}%
\end{equation}

An important characteristic for this cosmological model is that the field
equations possess a minisuperspace description. This means that the field
equations (\ref{eq.11})-(\ref{eq.14}) can be seen as the Euler-Lagrange
equations for the point-like Lagrangian function \cite{Paliathanasis:2023pqp}
\begin{equation}
L\left(  N,a,\dot{a},\phi,\dot{\phi},\dot{\psi}\right)  =-\frac{3}{N}\phi
a\dot{a}^{2}-\frac{3}{2N}a^{3}\dot{\phi}\dot{\psi}+\frac{N}{2}a^{3}V\left(
\phi\right)  .\label{eq.15}%
\end{equation}

From the modified Friedmann equations\ (\ref{eq.11}), (\ref{eq.12}) we can
define the effective energy density $\rho_{eff}$ and pressure components
$p_{eff}$ of $f\left(  Q\right)  $-gravity as follows%
\begin{align}
\rho_{eff}\left(  \Gamma^{B}\right)   &  =-\left(  \frac{3}{2N^{2}}\frac
{\dot{\phi}}{\phi}\dot{\psi}+\frac{V\left(  \phi\right)  }{2\phi}\right)
,~\label{eq.16}\\
p_{eff}\left(  \Gamma^{B}\right)   &  =-\frac{3}{2N^{2}}\frac{\dot{\phi}}%
{\phi}\dot{\psi}+\frac{V\left(  \phi\right)  }{2\phi}+\frac{2}{N}H\frac
{\dot{\phi}}{\phi}. \label{eq.17}%
\end{align}
Hence, the effective equation of state parameter reads%
\begin{equation}
w_{eff}\left(  \Gamma^{B}\right)  =\frac{p_{eff}\left(  \Gamma^{B}\right)
}{\rho_{eff}\left(  \Gamma^{B}\right)  }=1-\frac{2N\left(  NV\left(
\phi\right)  +2H\dot{\phi}\right)  }{N^{2}V\left(  \phi\right)  +3\dot{\phi
}\dot{\psi}}. \label{eq.18}%
\end{equation}

\section{Reconstruct cosmological solutions}

\label{sec3}

In this Section we reconstruct the scalar field potential $V\left(
\phi\right)  $, such that the field equations (\ref{eq.11})-(\ref{eq.14}) to
admit exact cosmological solutions of special interest. In this work we shall
extend our analysis within the case of other solutions of special
interests.\ Without loss of generality we assume that $a=e^{t}$ such that the
$H=\frac{1}{N}$, and $N$ is now the unknown function. Thus, the fluid
components for the cosmological fluid read
\begin{equation}
\rho_{eff}=\frac{3}{N^{2}},~p_{eff}=\left(  \frac{2}{3}\ln\left(  N\right)
^{\cdot}-1\right)  \rho_{eff} \label{eq.19}%
\end{equation}
while the equation of state is expressed%
\begin{equation}
w_{eff}=-1+\frac{2}{3}\ln\left(  N\right)  ^{\cdot}. \label{eq.20}%
\end{equation}

For connection $\Gamma^{B}$, \ from the field equations for the scalar fields
we derive the expressions%
\begin{align}
\phi\left(  t\right)   &  =\phi_{0}+\int^{t}e^{-3\tau}N\left(  \tau\right)
d\tau,\label{eq.21}\\
\dot{\psi}\left(  t\right)   &  =\frac{2}{3}\left(  1-e^{3\tau}\frac{\dot{N}%
}{N^{2}}\phi\right)  \label{eq.22}%
\end{align}
and for the scalar field potential we calculate%
\begin{equation}
V\left(  \phi\left(  t\right)  \right)  =-\frac{6}{N^{2}}\left(  \phi+\frac
{1}{3}e^{-3t}N\left(  1-e^{3t}\frac{\dot{N}}{N^{2}}\phi\right)  \right)  .
\label{eq.23}%
\end{equation}

In the following lines, we consider special functional forms for the lapse
function $N\left(  t\right)  $ which describe cosmological solutions of
special interest. We reconstruct the scalar fields, the scalar field
potential, and we determine the function $f\left(  Q\right)  ~$analytical or
numerical, from the Clairaut equation $V\left(  \phi\right)  =\left(
f(Q)-Qf^{\prime}\left(  Q\right)  \right)  $. Equivalently from the
following expression $f\left(  Q\right)  =V\left(  \phi\left(  Q\right)
\right)  -\phi\left(  Q\right)  V_{,\phi}\left(  Q\right)  .$

\subsection{de Sitter solution}

The de Sitter spacetime is recovered when $N\left(  t\right)  =N_{0}$. Then
for connection $\Gamma^{B}~$we calculate
\begin{align*}
\phi\left(  t\right)   &  =\phi_{0}-\frac{N_{0}}{3}e^{-3t},~\\
\dot{\psi}\left(  t\right)   &  =\frac{2}{3},\\
V\left(  \phi\left(  t\right)  \right)   &  =-\frac{6}{N^{2}}\left(
\phi+\frac{1}{3}e^{-3t}N_{0}\right)  .
\end{align*}
Therefore, scalar field potential reads $V\left(  \phi\left(  t\right)
\right)  =-\frac{6\phi_{0}}{N_{0}^{2}}$. This is a particular case, where the
limit of the GR is recovered in which $Q=Q_{0}$. \ Thus, the Clairaut equation
provides arbitrary function $f\left(  Q\right)  ~$\cite{ndim}.

\subsection{Scaling solution}

For $N=e^{\left(  \frac{3}{2}\left(  w_{0}+1\right)  \right)  t}$, from
(\ref{eq.19}) it follows $p_{eff}=w_{0}\rho_{eff},$ which describes a scaling
solution, with $w_{0}\leq1$. For the scalar fields it follows%
\begin{align*}
\phi\left(  t\right)   &  =\phi_{0}-\frac{3}{7-2w_{0}}e^{-\left(
7-2w_{0}\right)  \frac{t}{3}},\\
\dot{\psi}\left(  t\right)   &  =\frac{6}{7-2w_{0}}-\frac{4}{9}\left(
1+w_{0}\right)  \phi_{0}e^{\left(  7-2w_{0}\right)  \frac{t}{3}}\\
V\left(  \phi\left(  t\right)  \right)   &  =-\frac{2}{3}\phi_{0}\left(
7-2w_{0}\right)  e^{-\frac{4}{3}\left(  1+w_{0}\right)  t}.
\end{align*}
Hence, we derive the power-law potential%
\[
V\left(  \phi\right)  =V_{0}\left(  w_{0},\phi_{0}\right)  \left(  \phi
-\phi_{0}\right)  ^{\frac{4\left(  1+w_{0}\right)  }{7-2w_{0}}}.
\]
From the Clairaut equation we determine the power-law $f\left(  Q\right)  $
function, $f\left(  Q\right)  \simeq Q^{\frac{4\left(  1+w_{0}\right)
}{3\left(  2w_{0}-1\right)  }}$ \cite{ndim1}.

\subsection{$\Lambda$CDM}

Consider now that $N\left(  t\right)  =\left(  H_{0}\sqrt{\left(
1-\Omega_{m0}\right)  +\Omega_{m0}e^{-3t}}\right)  ^{-1},$ in order the
cosmological model to describe the $\Lambda$CDM. 

Then for the scalar fields we calculate%
\begin{align*}
\phi\left(  t\right)   &  =\phi_{0}-\frac{2H_{0}}{3\Omega_{m0}}\sqrt{\left(
1-\Omega_{m0}\right)  +\Omega_{m0}e^{-3t}},\\
\dot{\psi}\left(  t\right)   &  =\frac{\frac{4}{3}\Omega_{m0}e^{-3t}-\left(
\frac{4}{3}\left(  \Omega_{m0}-1\right)  +\frac{\phi_{0}}{H_{0}}\Omega
_{m0}\sqrt{\left(  1-\Omega_{m0}\right)  +\Omega_{m0}e^{-3t}}\right)
}{\left(  1-\Omega_{m0}\right)  +\Omega_{m0}e^{-3t}},\\
V\left(  \phi\left(  t\right)  \right)   &  =\frac{3\phi_{0}\Omega_{0}\left(
\left(  2-e^{-3t}\right)  \Omega_{m0}-2\right)  +4H_{0}\left(  1-\Omega
_{m0}\right)  \sqrt{\left(  1-\Omega_{m0}\right)  +\Omega_{m0}e^{-3t}}}%
{H_{0}^{2}\Omega_{m0}}.
\end{align*}
Therefore,%
\begin{equation}
V\left(  \phi\right)  \simeq-\frac{3}{4H_{0}^{4}}\left(  4H_{0}^{2}\left(
2\phi-3\phi_{0}\right)  \left(  \Omega_{m0}-1\right)  +9\left(  \phi-\phi
_{0}\right)  \phi_{0}\Omega_{m0}^{2}\right)  .
\end{equation}
The resulting $f\left(  Q\right)  $-gravity is $f\left(  Q\right)  \simeq
Q+\alpha_{1}\left(  \phi_{0},\Omega_{m0}\right)  Q^{2}+\alpha_{2}\left(
\phi_{0},\Omega_{m0}\right)  $, such that in the case where $\phi_{0}=0$, it
follows $f\left(  Q\right)  \simeq Q+\alpha_{2}\left(  0,\Omega_{m0}\right)
$, which is the limit of STEGR. This result is in agreement with that
presented in \cite{saikatfQlcdm}.

\subsection{Chaplygin Gas}

Consider now the lapse function $N\left(  t\right)  =\left(  \left(
1-\Omega_{m0}\right)  +\Omega_{m0}e^{3\mu t}\right)  ^{\frac{\mu}{2}}$, which
describes the cosmological solution with a Chaplygin gas, that is, the
equation of state parameter reads $p_{eff}=3^{-\mu}\left(  \Omega
_{m0}-1\right)  \rho_{eff}^{1+\mu}$.

For the scalar fields we calculate the analytic solution%
\begin{align*}
\phi\left(  t\right)   &  =\phi_{0}-\frac{e^{-3t}\left(  \left(  1-\Omega
_{m0}\right)  +\Omega_{m0}e^{3\mu t}\right)  ^{1+\frac{1}{2\mu}}}{3\left(
1-\Omega_{m0}\right)  }~_{2}F_{1}\left(  1,1-\frac{1}{2\mu},1-\frac{1}{\mu
},-\frac{\Omega_{m0}}{1-\Omega_{m0}}e^{3t}\right)  ,\\
\dot{\psi}\left(  t\right)   &  =\frac{2}{3}-\phi_{0}\Omega_{m0}e^{3\left(
1+\mu\right)  t}\left(  \left(  1-\Omega_{m0}\right)  +\Omega_{m0}e^{3\mu
t}\right)  ^{-1-\frac{1}{2\mu}}+\frac{_{2}F_{1}\left(  1,1-\frac{1}{2\mu
},1-\frac{1}{\mu},-\frac{\Omega_{m0}}{1-\Omega_{m0}}e^{3t}\right)  }{3\left(
1-\Omega_{m0}\right)  e^{-3\mu t}},\\
V\left(  \phi\left(  t\right)  \right)   &  =-2e^{-3t}\left(  \left(
1-\Omega_{m0}\right)  +\Omega_{m0}e^{3\mu t}\right)  ^{-\frac{1}{2\mu}%
}+\left(  6+3\Omega_{m0}\left(  e^{3\mu t}-2\right)  \right)  \phi.
\end{align*}
in which $_{2}F_{1}$ is the hypergeometric function.

\begin{figure}[ptbh]
\centering\includegraphics[width=0.9\textwidth]{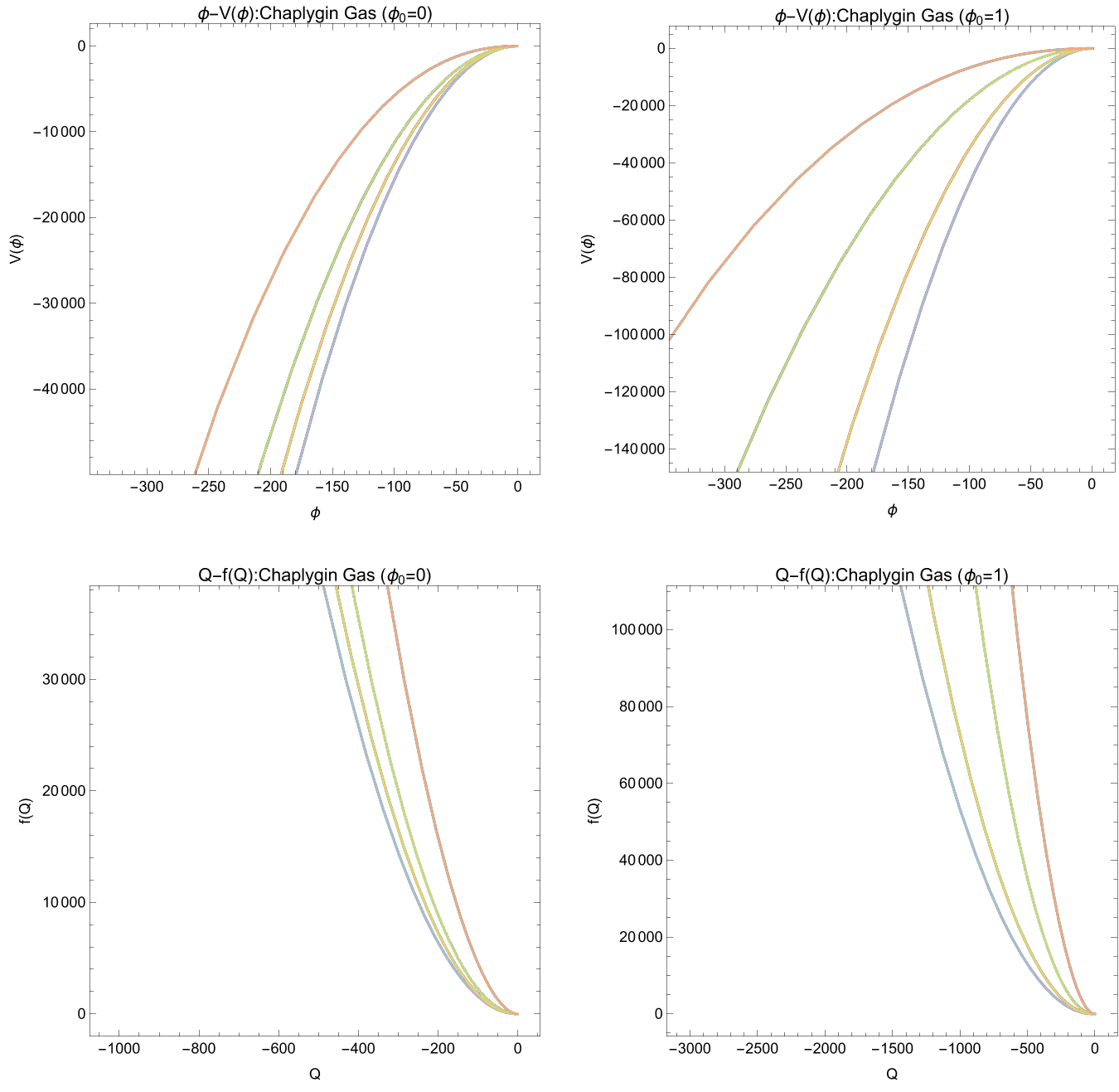}\caption{Chaplygin gas:
Scalar field potential $V\left(  \phi\right)  $ and $f\left(  Q\right)  $
function for the analytic solution of the Chaplygin gas for different values
of parameter $\mu$. Blue $\left(  \mu=-3\right)  $, orange $(\mu=-2),$ green
$(\mu=-1)$, red ($\mu=-\frac{1}{2}$). For the plot we considered $\Omega
_{m0}=0.3$. }%
\label{fig1}%
\end{figure}

Due to the nonlinearity of the solution, we can not present the closed-form
solution for the scalar field potential $V\left(  \phi\right)  $ in terms of
the scalar field $\phi$, or the corresponding function $f\left(  Q\right)  $.
Thus, in Fig. \ref{fig1} we present the parametric plots $\phi-V\left(
\phi\right)  $, and $Q-f\left(  Q\right)  $ for this analytic solution.

\subsection{Generalized Chaplygin Gas}

Consider now the lapse function $N\left(  t\right)  =\left(  \left(
1-\Omega_{m0}\right)  +\Omega_{m0}t\right)  ^{\frac{\mu}{2}}$, which describes
the cosmological solution with a Chaplygin gas, that is, the equation of state
parameter reads $p_{eff}=3^{-1-\mu}\Omega_{m0}\rho_{eff}^{1+\mu}-\rho_{m}$,
which is that of generalized Chaplygin gas. This model is known as
intermediate inflation.

For the scalar fields we calculate the analytic solution reads%
\begin{align*}
\phi\left(  t\right)   &  =\phi_{0}-\frac{e^{-3+\frac{3}{\Omega_{m0}}}\left(
1-\Omega_{m0}+\Omega_{m0}t\right)  ^{1+\frac{1}{2\mu}}}{\Omega_{m0}%
}\operatorname{Ei}_{\left(  -\frac{1}{2\mu}\right)  }\left(  3\left(
t+\frac{1}{\Omega_{m0}}-1\right)  \right)  ,\\
\dot{\psi}\left(  t\right)   &  =\frac{2}{3}-\frac{3e^{3t}\phi_{0}\Omega
_{m0}\left(  1-\Omega_{m0}+\Omega_{m0}t\right)  ^{-\frac{\mu}{2}}-3e^{3\left(
t+\frac{1}{\Omega_{m0}}-1\right)  }\left(  1-\Omega_{m0}+\Omega_{m0}t\right)
\operatorname{Ei}_{\left(  -\frac{1}{2\mu}\right)  }\left(  3\left(
t+\frac{1}{\Omega_{m0}}-1\right)  \right)  }{9\mu\left(  \left(  1-\Omega
_{m0}+\Omega_{m0}t\right)  \right)  },\\
V\left(  \phi\left(  t\right)  \right)   &  =-2e^{-3t}\left(  1-\Omega
_{m0}+\Omega_{m0}t\right)  ^{-\frac{\mu}{2}}+\frac{\phi}{\mu}\left(
1-\Omega_{m0}+\Omega_{m0}t\right)  ^{1+\frac{1}{2\mu}}\left(  \Omega_{m0}%
-6\mu\left(  1-\Omega_{m0}+\Omega_{m0}t\right)  \right)  ,
\end{align*}
where $\operatorname{Ei}$ denotes the exponential integral function.

\begin{figure}[ptbh]
\centering\includegraphics[width=0.9\textwidth]{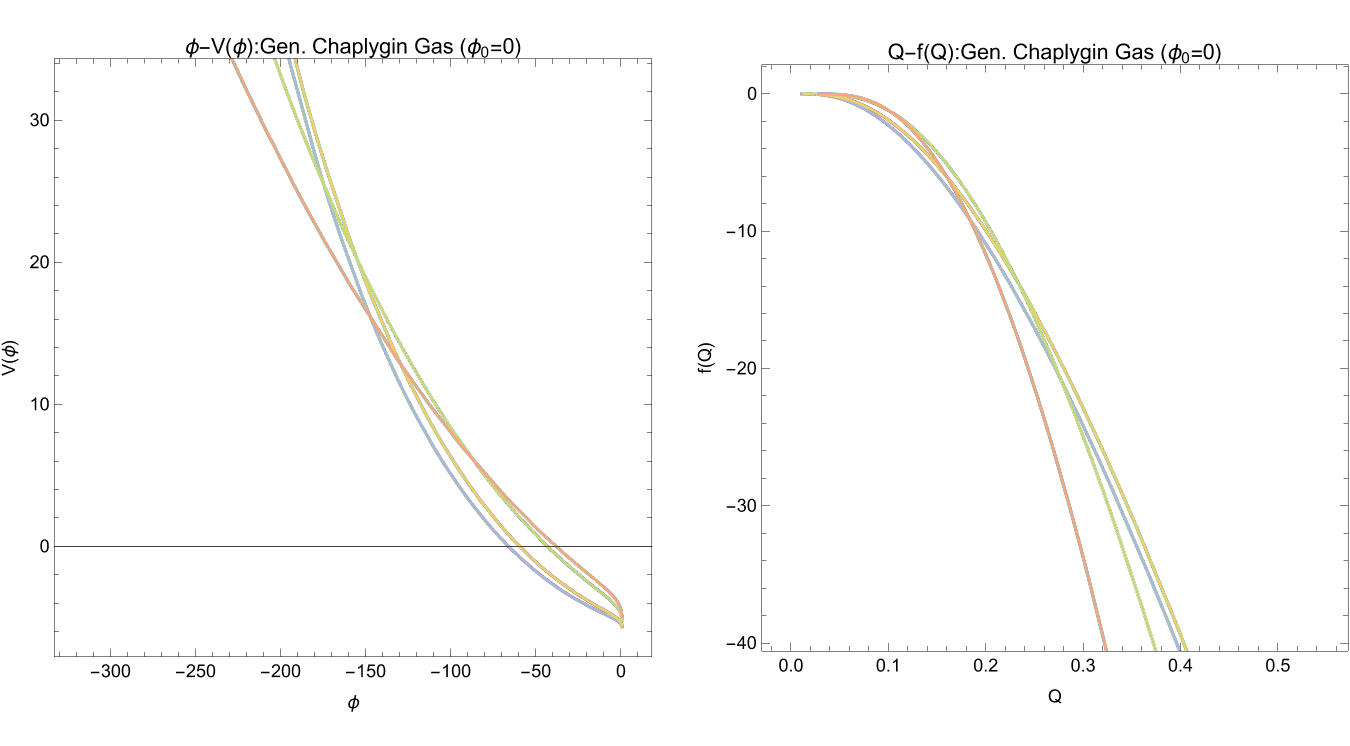}\caption{Generalized
Chaplygin gas: Scalar field potential $V\left(  \phi\right)  $ and $f\left(
Q\right)  $ function for the analytic solution of the generalized Chaplygin
gas for different values of parameter $\mu$. Blue $\left(  \mu=-5\right)  $,
orange $(\mu=-4),$ green $(\mu=-2)$, red ($\mu=-\frac{3}{2}$). For the plot we
considered $\Omega_{m0}=0.3$. }%
\label{fig2}%
\end{figure}

The resulting $V\left(  \phi\right)  $ and $f\left(  Q\right)  $ functions are
presented in Fig. \ref{fig2}. 

\subsection{CPL EoS}

The lapse function $N=\exp\left(  \frac{3}{2}\left(  w_{0}t+\frac{t^{2}}%
{2}w_{1}\right)  \right)  $ lead to the equation of state parameter
$w_{eff}=w_{0}+w_{1}\left(  1-\ln t\right)  $, that is, $w_{eff}=w_{0}%
+w_{1}\left(  1-a\right)  $, which that of the CPL model. For this model the
scalar fields and the potential function read%
\begin{align*}
\phi\left(  t\right)   &  =\phi_{0}+\frac{e^{-\frac{3\left(  w_{0}-2\right)
^{2}}{4w_{1}}}}{\sqrt{w_{1}}}\sqrt{\frac{\pi}{3}}\operatorname{erf}\left(
\frac{\sqrt{3}}{2\sqrt{w_{1}}}\left(  w_{1}t+w_{0}-2\right)  \right)  ,\\
\dot{\psi}\left(  t\right)   &  =\frac{2}{3}-\phi_{0}\left(  w_{0}%
+tw_{1}\right)  e^{-\frac{3t\left(  2\left(  w_{0}-2\right)  +tw_{1}\right)
}{4}}+\frac{e^{-\frac{3\left(  w_{0}-2+w_{1}\right)  ^{2}}{4w_{1}}}}%
{\sqrt{w_{1}}}\sqrt{\frac{\pi}{3}}\operatorname{erf}\left(  \frac{\sqrt{3}%
}{2\sqrt{w_{1}}}\left(  w_{1}t+w_{0}-2\right)  \right) \\
V\left(  \phi\left(  t\right)  \right)   &  =2e^{-\frac{3t\left(  2\left(
w_{0}-2\right)  +tw_{1}\right)  }{4}}+3\phi_{0}\left(  w_{0}+tw_{1}\right)
e^{-\frac{3t\left(  2w_{0}+tw_{1}\right)  }{2}}-\sqrt{\frac{3\pi}{w_{1}}%
}\left(  e^{-\frac{3}{4}\left(  4w_{0}t+2w_{1}t^{2}+\frac{\left(
w_{0}-2\right)  ^{2}}{w_{1}}\right)  }\right)  \operatorname{erf}\left(
\frac{\sqrt{3}}{2\sqrt{w_{1}}}\left(  w_{1}t+w_{0}-2\right)  \right)  ,
\end{align*}
where now where $\operatorname{erf}$ is the error function.

\begin{figure}[ptbh]
\centering\includegraphics[width=0.9\textwidth]{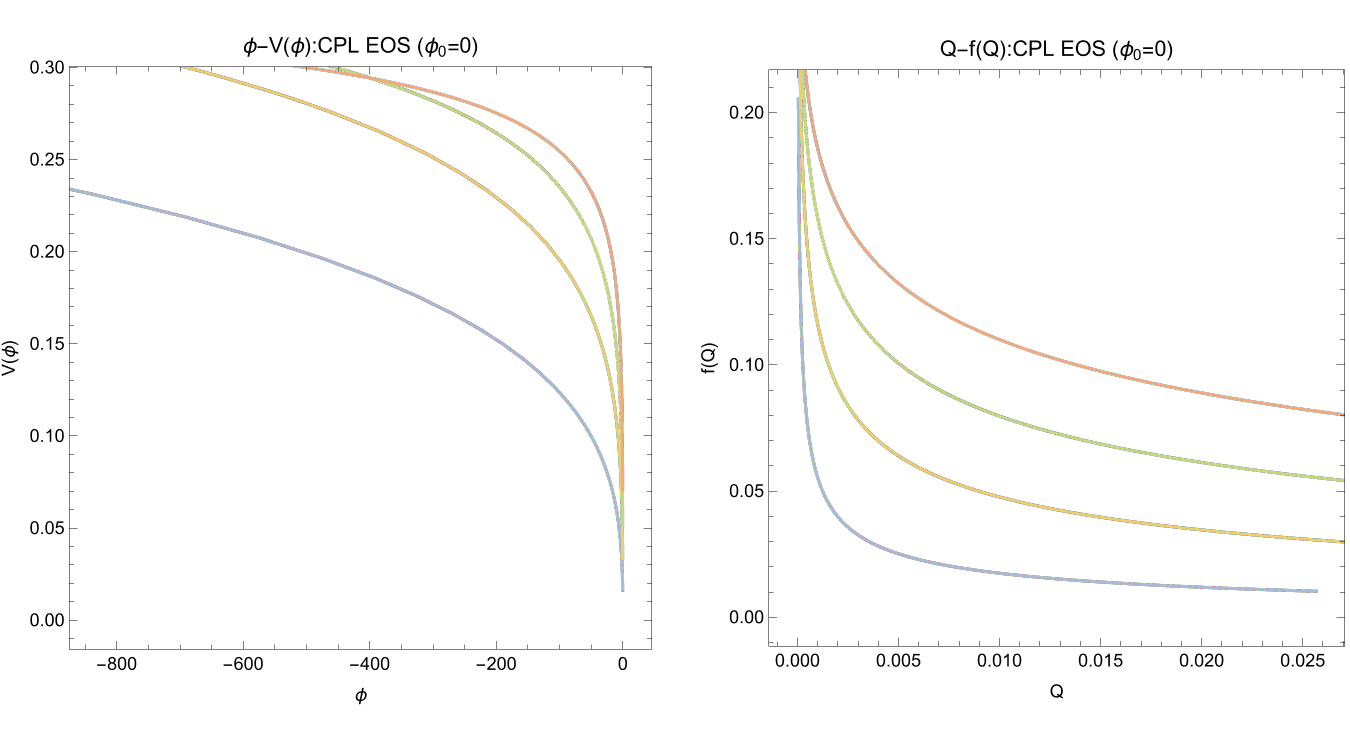}\caption{CPL EoS:
Scalar field potential $V\left(  \phi\right)  $ and $f\left(  Q\right)  $
function for the analytic solution of the CPL\ EoS for different values of
parameter $w_{1}$. Blue $\left(  w_{1}=0.1\right)  $, orange $(w_{1}=0.2),$
green $(w_{1}=0.3)$, red ($w_{1}=0.4$). For the plot we considered
$w_{1}=-0.9$. }%
\label{fig3}%
\end{figure}

The resulting functions $V\left(  \phi\right)$ and $f\left(  Q\right) $ are
presented in Fig. \ref{fig3}. 

\section{Conclusions}

\label{sec4}

In this work we studied exact cosmological solutions in the framework of non-coincident $f(Q)$ gravity. 
Focusing on the connection branch $\Gamma_B$, we reformulated the field equations in a scalar-tensor representation with a minisuperspace Lagrangian, which facilitated the derivation and classification of exact solutions. We showed that a wide range of cosmological scenarios can be accommodated, including the de Sitter universe, scaling solutions with constant equation of state, $\Lambda$CDM-type behavior, and Chaplygin gas models (both standard and generalized), together with CPL-type parameterizations. 

The explicit reconstruction of the scalar field potential and the corresponding $f(Q)$ function demonstrated the integrability of these models in closed or parametric form. Importantly, our results confirm that non-coincident formulations of $f(Q)$ gravity possess a richer solution space than the coincident gauge, allowing for consistent realizations of both inflationary and dark energy eras within a unified geometric framework. 

Future research directions include the stability analysis of the reconstructed solutions, their confrontation with precision cosmological data, and extensions to anisotropic or inhomogeneous backgrounds. These steps will be essential for assessing the viability of non-coincident $f(Q)$ gravity as a compelling alternative to $\Lambda$CDM and other modified gravity scenarios.

\end{document}